\documentclass[a4paper]{article}
\makeatletter
\def\@seccntformat#1{\@ifundefined{#1@cntformat}%
   {\csname the#1\endcsname\quad}  
   {\csname #1@cntformat\endcsname}
}
\let\oldappendix\appendix 
\renewcommand\appendix{%
    \oldappendix
    \newcommand{\section@cntformat}{\appendixname~\thesection\quad}
}
\makeatother

\usepackage[dvipdfmx]{graphicx}

\usepackage{cite}
\usepackage[normalem]{ulem}
\usepackage{url}

\usepackage{amssymb,amsfonts,amsmath}
\usepackage[top=25truemm,bottom=25truemm,left=20truemm,right=20truemm]{geometry}
\usepackage{bm}
\usepackage{cases}
\usepackage[mathlines]{lineno}
\usepackage{multirow}
\usepackage{amssymb}

\title{Winning by hiding behind others: An analysis of speed skating data}
\bigskip
\author{Genki Ichinose${}^{1}$, Daiki Miyagawa${}^{1}$, Junji Ito${}^{2}$, Naoki Masuda${}^{3,4,5*}$
\ \\
\ \\
${}^{1}$
{\it Department of Mathematical and Systems Engineering, Shizuoka University, }\\
{\it 3-5-1 Johoku, Naka-ku, Hamamatsu, 432-8561, Japan}\\
${}^{2}$
{\it Skating Club, Yamanashi Gakuin University, 2-4-5 Sakaori, Kofu, Yamanashi, 400-8575, Japan}\\
${}^{3}$
{\it Department of Mathematics University at Buffalo, }\\
{\it State University of New York, Buffalo, NY 14260-2900, USA}\\
${}^{4}$
{\it Computational and Data-Enabled Science and Engineering Program, }\\
{\it University at Buffalo, State University of New York, Buffalo, NY 14260-5030, USA}\\
${}^{5}$
{\it Faculty of Science and Engineering, Waseda University,}\\
{\it 3-4-1 Okubo, Shinjuku-ku, Tokyo 169-8555, Japan}\\
${}^{*}$
Corresponding author (naokimas@buffalo.edu)
}

\begin{document}

\maketitle

\section*{Abstract}
In some athletic races, such as cycling and types of speed skating races, athletes have to complete a relatively long distance at a high speed in the presence of direct opponents. To win such a race, athletes are motivated to hide behind others to suppress energy consumption before a final moment of the race. This situation seems to produce a social dilemma: players want to hide behind others, whereas if a group of players attempts to do so, they may all lose to other players that overtake them. To support that speed skaters are involved in such a social dilemma, we analyzed video footage data for 14 mass start skating races to find that skaters that hid behind others to avoid air resistance for a long time before the final lap tended to win. Furthermore, the finish rank of the skaters in mass start races was independent of the record of the same skaters in time-trial races measured in the absence of direct opponents. The results suggest that how to strategically cope with a \textit{skater's dilemma} may be a key determinant for winning long-distance and high-speed races with direct opponents.

\section*{Keywords}
Mass start speed skating, chicken game, game theory

\section{Introduction}

Pacing behavior and decision making on it have been studied in various athletic races \cite{Abbiss2008SportsMed, Thompson2014book, Smits2014SportsMed, Renfree2014SportsMed}.
Experimental \cite{Hettinga2011BrJSportsMed, Hettinga2012BrJSportsMed} and modeling \cite{deKoning1992JBiomech, vanIngenSchenau1990JBiomech, vanIngenSchenau1990MedSciSportsExerc, deKoning1999JSciMedSport} studies as well as analyses of real races \cite{Muehlbauer2010ResQExercSport, Muehlbauer2010IntJSportsPhysiolPerform} revealed strategic benefits of pacing behavior. Earlier studies investigated pacing behavior in time-trial races in which interaction between athletes is not considered to be a dominant factor. Examples include long-track speed skating \cite{Muehlbauer2010ResQExercSport, Muehlbauer2010IntJSportsPhysiolPerform}, swimming \cite{Mauger2012MedSciSportsExerc}, and some types of cycling races \cite{Corbett2009IntJSportsPhysiolPerform}.

Pacing behavior also abounds in races with direct opponents, such as short-track speed skating \cite{Bullock2008IntJPerformAnalSport, Muehlbauer2011EurJSportSci, Konings2016IntJSportsPhysiolPerform, Noorbergen2016IntJSportsPhysiolPerform, Menting2019IntJSportsPhysiolPerform} and marathon \cite{Ely2008MedSciSportsExerc,Renfree2013IntJSportsPhysiolPerform,Angus2014JSportsSci}. For example, in a marathon race, the top runners were faster than the other runners in every 5 km segment and their speeds were kept less variable over the race than the other runners' speeds \cite{Renfree2013IntJSportsPhysiolPerform}.
In addition, tactical positioning has also been analyzed in races with direct opponents such as short-track speed skating \cite{Maw2006JSportsSci, Bullock2008IntJPerformAnalSport, Muehlbauer2011EurJSportSci, Haug2015IntJSportsPhysiolPerform, Konings2016IntJSportsPhysiolPerform, Noorbergen2016IntJSportsPhysiolPerform, Konings2018SportsMed, Menting2019IntJSportsPhysiolPerform} and cycling \cite{Moffatt2014JQuantAnalSports}. For example, in 500-m short-track speed skating, where approximately four skaters compete in each race, being positioned at an inner side of the oval at the start of a race \cite{Noorbergen2016IntJSportsPhysiolPerform, Maw2006JSportsSci, Muehlbauer2011EurJSportSci} and securing the first rank at the beginning of a race \cite{Noorbergen2016IntJSportsPhysiolPerform, Haug2015IntJSportsPhysiolPerform} tend to increase the probability to be top finishers.
Pacing behavior and tactical positioning are not distinct strategies. For example,
in 1000 m and 1500 m short-track speed skating races, skaters that gradually speed up and improve the rank towards the finish are more likely to win than those who speed high in the beginning of the race \cite{Konings2016IntJSportsPhysiolPerform, Noorbergen2016IntJSportsPhysiolPerform}.

In pacing, an athlete's strategic thinking and the implemented action are relatively, though not exclusively, independent of other athletes' choices. In contrast, in the case of the competition for an inner-side spot at the start of the race, athletes explicitly have to compete with each other by taking into account how other athletes would behave, because there is less such spots than the number of athletes. The difference between these two types of strategic situations may be mapped to the difference between a single-person and multi-person games. Both components are probably indispensable for athletes. Here we focus on athletes' strategic actions in essentially multi-person situations.

In long-distance short-track speed skating races, not leading the race from the beginning has been shown to be a sensible strategy presumably because leading a race would cost energy to skaters. In some types of races such as speed skating and cycling, air resistance is strong due to the high speed of athletes' movements and therefore is considered to be
a key determinant of the race outcome \cite{Thompson2014book}.
In these types of races, athletes are generally motivated to hide behind others to avoid attrition due to air resistance. They may also be motivated to avoid being followed closely by others to be exploited as a shield.
In fact, unless the race's speed is low, at least one athlete has to receive a high air resistance at any moment of time. Therefore,
we propose that air resistance creates a multi-person game of a social dilemma type \cite{Sugden1986book, Fudenberg1991book, Osborne1994book, Nowak2006book}. In other words, we propose that players (i.e., athletes) want to use others as a shield, and if a group of athletes tries to do so, the entire group would be slowed down and superseded by other players.

To investigate such a social dilemma scenario, we focus on speed skating and test our hypothesis that
the skaters who hide behind others before a final part the race tend to win, while those who lead the race in those laps would lose.
We examine data of mass start skating races, which are a relatively new type of race that appeared in the 2018 Winter Olympics for the first time. Furthermore, we show that the success of skaters in mass start races is more strongly related to the time for which they hide behind other skaters than to their performance in time-trial races without evident competitors.

\section{Methods}

\subsection{Data}

A mass start race consists of 16 laps around a 400-m oval track, totaling a distance of 6400-m.
A race accommodates up to 24 skaters that simultaneously start on the same start line.
In the World Cup, of which the data we use, the first, second, and third finishers earn 60, 40, and 20 points, respectively.
In addition, skaters who passed the finish line completing the fourth, eighth, and twelfth laps are awarded with a premium of
5, 3, and 1 points, respectively, for each of the three intermediate laps.
We refer to the order in which the skater passed the finish line in the final lap as the finish rank.
We use this term to distinguish it from the rank based on the total points that the skaters earn in the race, which we refer to as the final rank.
By definition, a small rank value corresponds to a good performance, such as finishing early or obtaining a high score.
Because the points given to the first three finishers are much larger than the premiums given at intermediate laps, the top three skaters are unchanged regardless of whether we use the finish rank or final rank. We focus on the finish rank in the present study because we analyze the data in the last few laps and aim to shed light on strategies used by skaters aiming to finish in top three.

We obtained the race results from the official website of the International Skating Union (ISU) \cite{ISUresult} and collected the video footage from YouTube \cite{YouTube}.
We do not need ethics approval because all data are publicly available.
We used the data obtained from the eight races in ISU World Cups in the 2016-2017 season and the six races in the 2017-2018 season. We excluded the last World Cup for men in the 2017-2018 season held in Minsk, Republic of Belarus, because the number of competitors (i.e., 8) was much smaller than in the other 15 races (i.e., at least 13).
We also excluded the same World Cup in Minsk for women 
because the video footage was not available on YouTube.

In the remaining races, we observed some disqualified skaters.
In mass start races, a skater who is one lap behind the fastest skater becomes disqualified.
We assigned the worst finish rank to the disqualified skater.
If there were multiple disqualified skaters in a race, a disqualified skater that was overtaken by the fastest one earlier was assigned with a worse finish rank than those overtaken later.
Moreover, there was one skater who was disqualified presumably because of excessive offending behavior to a different skater in the World Cup for women in the 2017-2018 season held in Heerenveen, the Netherlands. We assigned the worst finish rank to this skater.

We also tested whether strong skaters in time-trial races are strong in mass start races. 
Therefore, for each skater, we obtained the best time among the time-trial races in 2016/2017 and 2017/2018 seasons. To this end, we used seven World Cup 3000-m (Division A) races for women and six World Cup 5000-m (Division A) races for men because these distances are the closest to that of mass start races, i.e., 6400-m.

\subsection{Intermediate ranks}

We computed two types of quantities from the video footage.
First, we calculated for each race the intermediate ranks of the skaters at the last 3, 2, and 1 laps. We only computed these intermediate ranks for the top three finishers because our primary interests are the location of top finishers during the race and whether they successfully avoid air resistance by positioning behind other competitors.
Two checkers (authors GI and DM) independently recorded each skater's intermediate rank at the last 3, 2, and 1 laps. If the two checkers agreed on the rank, we adopted it. Otherwise, we set the intermediate rank of the skater at a given lap to the average over the two checkers.

\subsection{Time exposed to high air resistance}

Second, we calculated the total time in seconds for which each skater headed a group to presumably receive a high air resistance, which we call the exposed time. We specifically defined the exposed time, denoted by $\tau$, of a skater as the length of time for which the skater leads a group between when the first skater passed the finish line at the last three laps and when the first skater passed the finish line at the last one lap. If a skater X is behind a different skater Y's back within a certain distance, as we will detail below, we regard that X is not leading a group. If X is not behind any other skater, X is defined to be leading a group.

Skater X is defined to be behind skater Y if X's entire head falls within the range of the shoulder width of skater Y along the horizontal direction. Because we could not directly measure the distance from the video footage, we measured the time difference between skaters X and Y to judge whether or not X is sufficiently closely following Y to be able to use Y or other skaters as a shield. To this end, we recorded the time difference between the two skaters at the four boundary locations between a straight and a curve in the 400-m oval. At each of the four boundaries, we measured the time difference  between each pair of skaters X and Y for which X's entire head was included in the shoulder width of Y. If and only if the time difference was larger than 0.2 seconds, which account for approximately 2.5 m, we regarded that the second skater X was leading a group or in a solo, receiving a high air resistance. In this case, we added the time that X spent to proceed from this boundary to the next boundary to X's $\tau$.

To reliably determine $\tau$, the two checkers independently watched the video to find out when each skater passed each boundary line.
Let $\tilde{\tau}_i$ ($i=1, 2$) be the exposed time observed by checker $i$.
When the relative discrepancy defined by $D=|\tilde{\tau}_1-\tilde{\tau}_2|/(\tilde{\tau}_1+\tilde{\tau}_2)$ is smaller than $0.1$, then we considered that the two checkers sufficiently agreed on the value of the exposed time, and we used 
the average $(\tilde{\tau}_1+\tilde{\tau}_2)/2$ as $\tau$.
Otherwise, the two checkers independently measured the time again until $D$ is less than $0.1$. Once $D<0.1$ was satisfied, we set $\tau = (\tilde{\tau}_1+\tilde{\tau}_2)/2$.

In each race, skaters sometimes disappeared from the video footage, which occurred when they were out of the camera's view.
In this case, we regarded that the skaters kept the preceding state, where the state refers to either the rank, which was used in the calculation of the intermediate rank, or the exposition (or the lack thereof) to the high air resistance, which was used in the calculation of $\tau$.
When the skaters that had disappeared re-appeared in the footage, we checked the state of the skaters again and reset it to the current state if it was different from the last observed state. During the third last lap to the last lap, the longest time for which the skaters continuously disappeared from the footage was 10.48 s on average. Because one lap needs 59.38 s on average (although this average depends on the sex among other things), the fraction of time for which the skaters disappear from the footage in these laps, which should affect the accuracy of estimating $\tau$, is equal to $10.48/(2\times 59.38) = 0.088$ on average.

\subsection{Exclusion of races in which some skaters broke away early}

We found that there were two broad types of races. In one type, some skaters attempted to break away relatively early in the race
and succeeded to finish in tops. In the other type of race, top finishers
positioned behind other skaters to save energy and then accelerated to overtake them somewhere near the finish. In the present study, we are only interested in the latter type of race. 

We determined the type of each race as follows. We inspected the video footage between the time when the first skater passed the finish line at the second last lap and the time when the first skater (who may be a different skater) passed the finish line at the last lap. If the minimum time distance between the $n$th and $(n+1)$th skaters, where $n=1, 2$, or 3, was more than two seconds for any $n$, we classified the race to be of breaking-away type. With a two second difference between the $n$th and $(n+1)$th skaters in the second last to the last lap, it is practically impossible for the $(n+1)$th and following skaters to overtake the preceding skaters. The two checkers independently checked all the races according to these criteria. They both classified the same five races out of the 14 races as the breaking-away type. Therefore, in the following analysis, we focus on the remaining nine races.

\section{Results}

\begin{figure}[bt]
	\centering
	\includegraphics[width=4in]{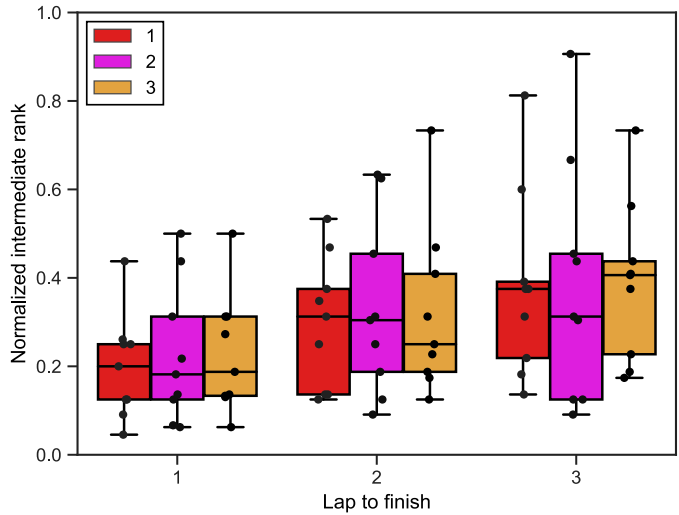}
	\caption{Relationships between the intermediate rank, finish rank, and laps to the finish, for the top three finishers. A circle represents the skater whose finish rank is indicated by the color of the bar in the background. We horizontally jittered the circles to avoid their overlapping.}
	\label{lap-IntRankAve}
\end{figure}

We first analyzed the relationship between the intermediate rank of skaters, the lap to the finish, and the finish rank for the top finishers. The intermediate rank as a function of the lap to the finish and the finish rank is shown in Fig.~\ref{lap-IntRankAve}. To compare across the different races, we employed the normalized intermediate rank, which is defined as the finish rank divided by the number of skaters in the race. Each filled circle in the figure represents a race. The figure suggests that the top three finishers tend to gradually improve their ranks towards the end of the race, which is consistent with the previous results for 1000 and 1500-m short-track speed skating \cite{Konings2016IntJSportsPhysiolPerform, Noorbergen2016IntJSportsPhysiolPerform, Menting2019IntJSportsPhysiolPerform, Bullock2008IntJPerformAnalSport}. The figure also suggests that the intermediate ranks are not apparently related to the finish rank for the top three finishers.

To statistically test this observation, we adopted a linear mixed model (LMM).
We incorporated the skater's ID as a random effect (i.e., random intercept) into the model because various skaters appear in multiple races and thus multiple times in the data.
We carried out the statistical analysis using R 3.6.3 with lme4 package. 
The LMM is given by
\begin{equation}
I_{i, j} = \beta_0 + \beta_{1}L_{i, j}+\beta_{2}F_{i, j}+S_i,
\end{equation}
where $i$ represents a skater's ID, $j$ represents a race's ID $(j=1,\ldots, 9)$. $I_{i, j}$ ($0< I_{i, j} \le 1$) is the normalized intermediate rank, $L_{i, j}$ ($L_{i, j} = 1, 2$, or $3$) is the lap to finish, and $F_{i, j}$ ($F_{i, j} = 1, 2$, or $3$) is the finish rank of skater $i$ in race $j$. Variable $S_i$ represents the random effect (i.e., random intercept) nested within each skater.
We listed the dependent and explanatory variables for this and the following two LMMs in Table \ref{LMM-variable}.
We assumed that the random effect was normally distributed with mean zero. 
We found that the effect of the lap to the finish was significant ($\beta_{1}=0.0827$, 95\% confidential interval (CI): $[0.0439,0.1215]$, $p = 8.65 \times 10^{-5}$, $n=81$), whereas that of the finish rank was not
($\beta_{2}=0.0238$, CI: $[-0.0310,0.0785]$, $p = 0.403$, $n=81$). These results support the casual observation that we made with Fig.~\ref{lap-IntRankAve}.

\begin{table}[bt]
	\centering
	\caption{Variables used in the LMMs, i.e., Eqs.~(1)--(3).}
	\begin{tabular}{lll}
	\hline
	\multicolumn{3}{l}{Dependent variables}           \rule[-2mm]{0mm}{8mm}  \\
	  & \ \ $I_{i,j}$         &\ \ Normalized intermediate rank  \rule[0mm]{0mm}{5mm} \\
	  & \ \ $\tilde{F}_{i,j}$ &\ \ Normalized finish rank        \rule[0mm]{0mm}{5mm} \\
	\multicolumn{3}{l}{Explanatory variables}        \rule[-2mm]{0mm}{8mm}  \\
	  & \ \ $L_{i,j}$         &\ \ Lap to finish                 \rule[0mm]{0mm}{5mm} \\
	  & \ \ $F_{i,j}$         &\ \  Finish rank                   \rule[0mm]{0mm}{5mm} \\
	  & \ \ $\tau_{i,j}$      &\ \  Exposed time                  \rule[0mm]{0mm}{5mm} \\
	  & \ \ $T_i$             &\ \  Best time in time-trial races \rule[0mm]{0mm}{5mm} \\
	  & \ \  $S_i$             &\ \  Random effect                 \rule[0mm]{0mm}{5mm} \\ \hline
	\end{tabular}
	\label{LMM-variable}
\end{table}

\begin{figure}[bt]
	\centering
	\includegraphics[width=6in]{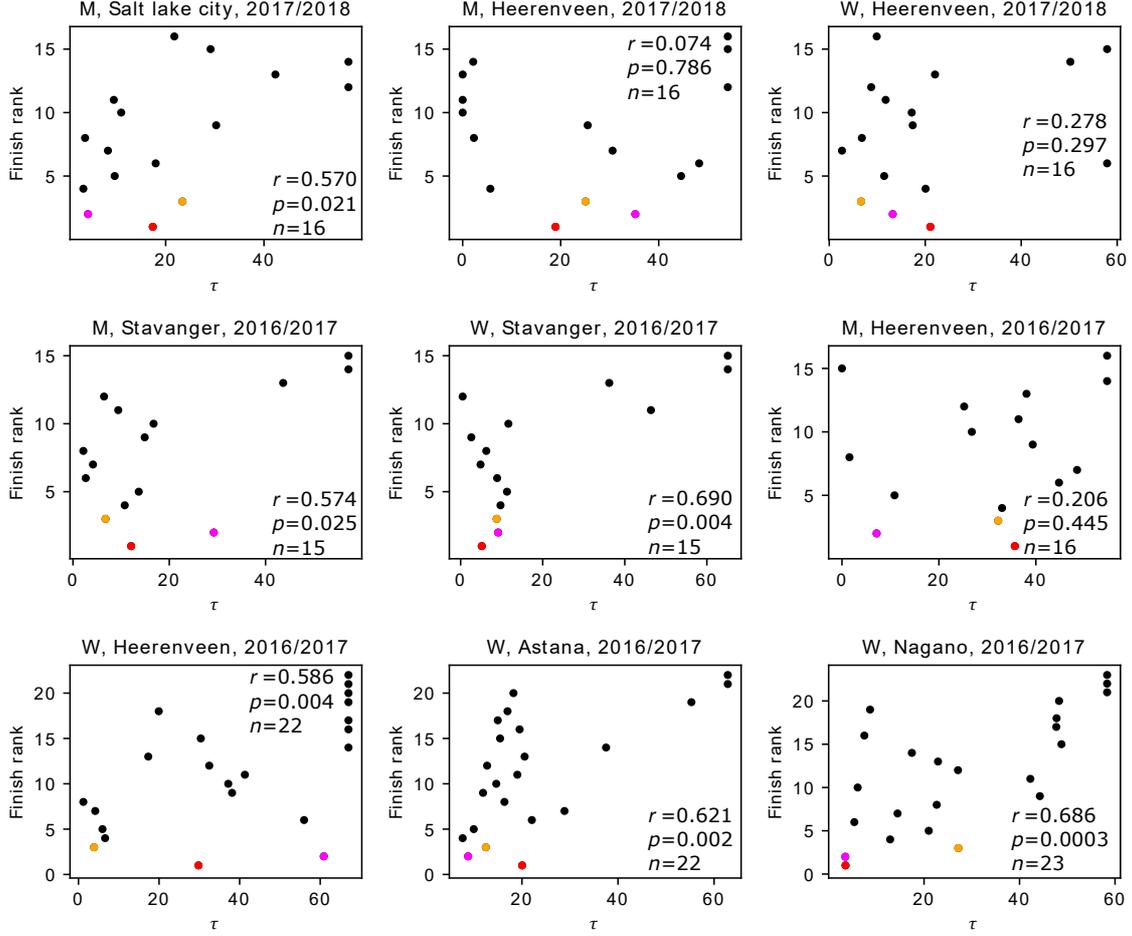}	
	\caption{Relationships between the finish rank and $\tau$. Each panel represents a race. A circle represents a skater. The top three finishers are shown in red, magenta, and orange. The Pearson correlation coefficient, $r$, its $p$ value, and the sample size (i.e., number of skaters in the race), $n$, are also shown. M: men. W: women.}
	\label{tA-finish}
\end{figure}

\begin{figure}[bt]
	\centering
	\includegraphics[width=4in]{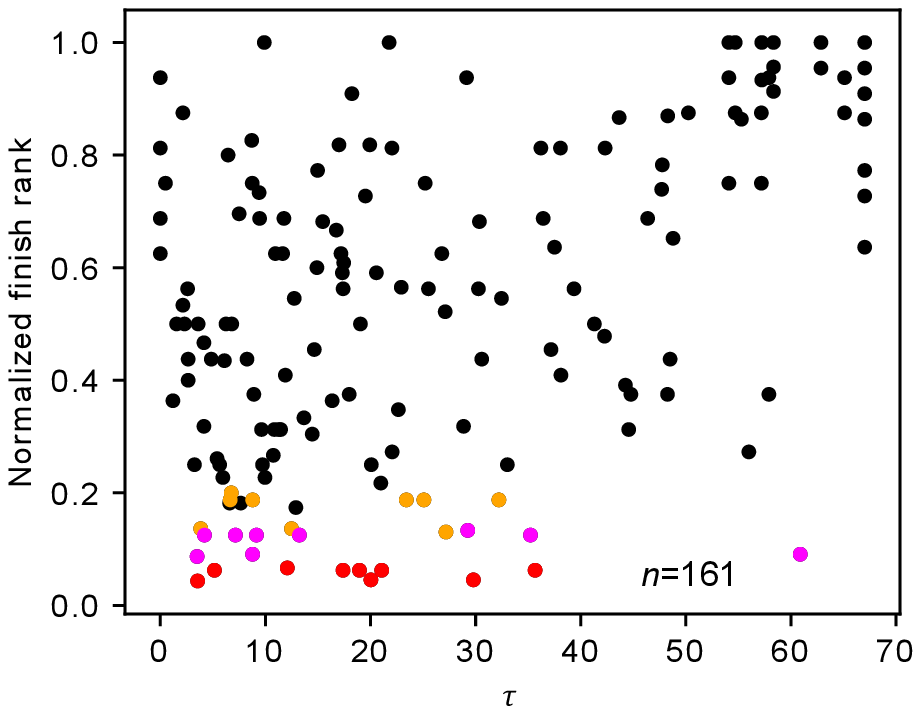}	
	\caption{Relationships between the normalized finish rank and $\tau$ superposed over all races. The top three skaters are shown in red, magenta, and orange.}
	\label{tA-finishNorm}
\end{figure}

The relationships between the finish rank and 
the exposed time, $\tau$ (i.e., the amount of time for which the skater leads a group of skaters between the third last to the last laps),
are shown in Fig.~\ref{tA-finish} for each race.
The finish rank and $\tau$ are positively correlated in all the races and significantly so in six out of the nine races despite relatively small sample sizes (i.e., number of skaters) in each race. Therefore, skaters with large $\tau$ tended to finish late, probably because such skaters were subject to high air resistance for a long time. An overlay of the results obtained from all the races using the normalized finish rank is shown in Fig.~\ref{tA-finishNorm}. This figure apparently confirms the tendency revealed for the individual races shown in Fig.~\ref{tA-finish}. We then fitted the following LMM to the data shown in
Fig.~\ref{tA-finishNorm}:
\begin{equation}
\tilde{F}_{i,j} = \beta_0 + \beta_1 \tau_{i,j} + S_i,
\end{equation}
where $i$ represents the skater's ID $(i=1,\ldots, 70)$, $j$ represents the race's ID $(j=1,\ldots, 9)$, and $\tilde{F}_{i,j}$ is the normalized finish rank of skater $i$ in race $j$. The influence of the exposed time on the normalized finish rank was significant ($\beta_{1}=0.0985$, CI: $[0.0669, 0.1312]$, $p = 6.25 \times 10^{-9}$, $n=161$).

\begin{figure}[bt]
	\centering
	\includegraphics[width=5.5in]{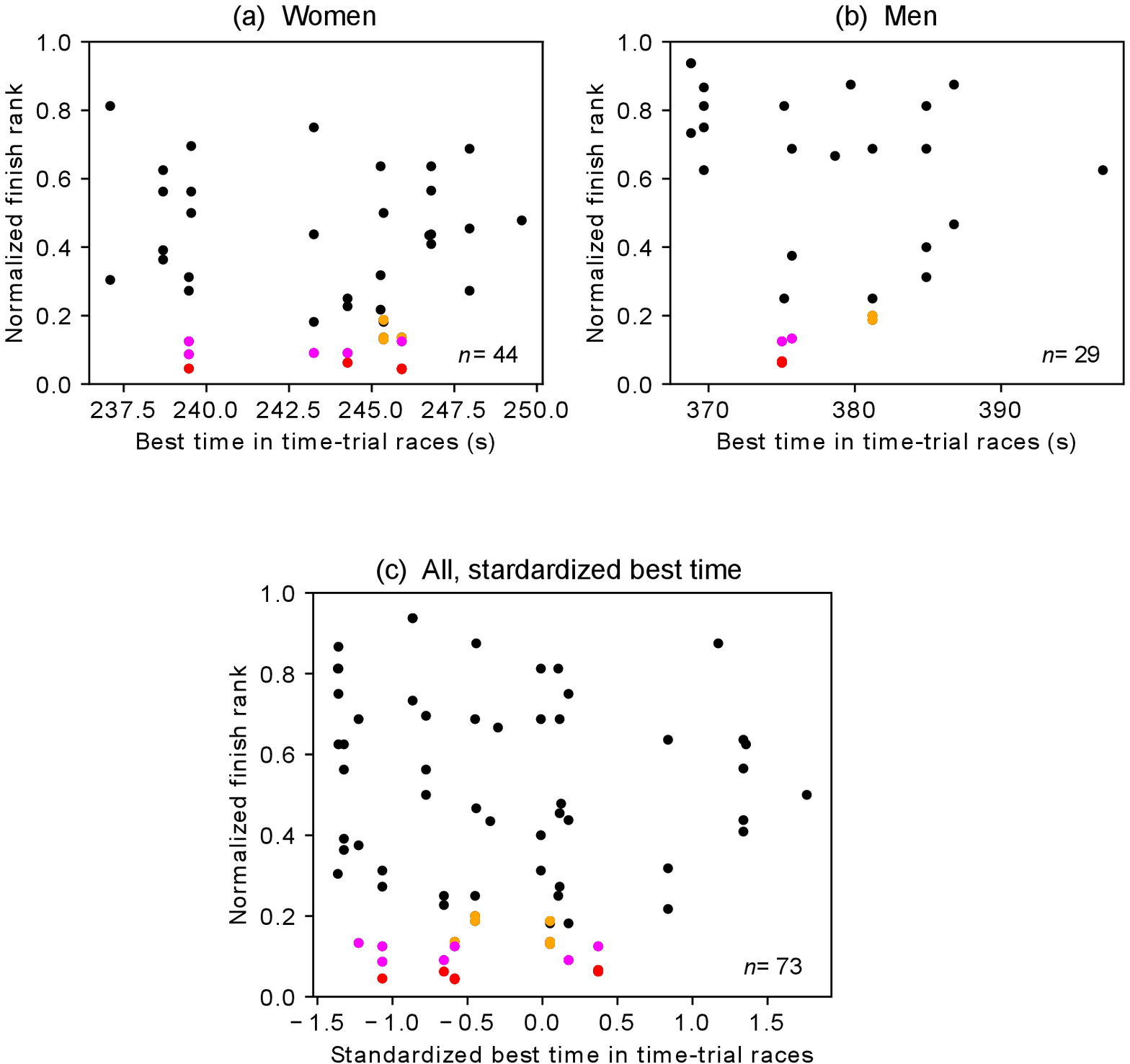}
	\caption{Relationships between the normalized finish rank and the best time in time-trial races. A circle represents a skater. The top three skaters in mass start races are shown in red, magenta, and orange. (a) Women. (b) Men. (c) Both sexes combined using the standardized best finish time.}
	\label{fig:time trials}
\end{figure}

To examine whether strong skaters in long-track speed skating are also strong in mass start races, we examined the relationship between the normalized finish rank and the best time for each skater in long-track speed skating. The best finish time for each skater is defined to be the one among
all the 3000-m or 5000-m long-track speed skating races
for women and men, respectively, over 2016/2017 and 2017/2018 seasons.
The relationships between the normalized finish rank and the best finish time 
are shown in Fig.~\ref{fig:time trials}(a) and \ref{fig:time trials}(b) for the women and men, respectively.
It should also be noted that there are less samples in Fig.~\ref{fig:time trials} than in Fig.~\ref{tA-finishNorm} because many skaters competed in mass start races and not in time-trial races in these two seasons; these skaters do not appear in Fig.~\ref{fig:time trials}.
We fitted the LMM given by
\begin{equation}
\tilde{F}_{i,j} = \beta_0 + \beta_1 T_i + S_i
\label{eq:LMM time-trial}
\end{equation}
to the relationship between the normalized finish rank of skater $i$ in mass start race $j$, $\tilde{F}_{i, j}$, and the best time in time-trial races for skater $i$, denoted by $T_i$, separately for each sex.
The influence of the best time in time-trial races on the normalized finish rank was not significant for either sex (women: $\beta_1=-0.0089$, CI: $[-0.4550, 0.4401]$, $p=0.970$, $n=44$; men: $\beta_1=-0.0361$, CI: $[-0.3677, 0.2997]$, $p=0.837$, $n=29$).

Because the finish time in long-track speed skating also heavily depends on conditions such as the altitude and indoor versus outdoor rinks, we also
examined the relationship between the normalized finish rank and a standardized best time for each skater in long-track speed skating.
We calculated the standardized finish time for each skater in a race as the $Z$ score of the finish time of the skater in the race (i.e., the actual finish time subtracted by the average for the race, which is then divided by the standard deviation for the race). 
The standardized best time for a skater is the lowest $Z$ score for the skater across all the 3000-m (for women) or 5000-m (for men) races in 2016/2017 and 2017/2018 seasons.
The relationships between the normalized finish rank and the standardized best time 
are shown in Fig.~\ref{fig:time trials}(c). With the LMM given by Eq.~\eqref{eq:LMM time-trial}, where we replaced $T_i$ by the standardized best time, we found that the influence of the standardized best time in time-trial races on the normalized finish rank was insignificant ($\beta_1=0.4076$, CI: $[-1.1781, 1.9999]$, $p=0.618$, $n=73$). These results suggest that the competence in time-trial races is not related to that in mass start races.

\section{Discussion}

We showed that the exposed time, i.e., the amount of time for which the skater directly receives a high air resistance without being located behind another skater, was negatively correlated with the finish rank in mass start races. On the other hand, the finish rank was uncorrelated with the skater's competence in time-trial races. These results suggest that, in mass start races, skills to use other skaters as a shield may be more important than how fast they can run. Possible generalization of the present results to the case of other types of races where direct opponents exist and the avoidance of air resistance seems to be a key factor, such as short-track speed skating and cycling, warrants future work.

The present results suggest that skaters in mass start races are motivated to avoid air resistance before the final portion of the race. However, if all skaters avoid leading a group, a group will not be formed. Then, the skaters may proceed extremely slowly, with every skater giving a way to others to wait until somebody starts to lead a group. In actual races, such an extreme slowing down would not happen, except in early laps, and one skater typically accepts to lead a linear group of skaters and receives a high air resistance, at least for some duration of time. If every skater in group G waits until somebody decides to lead G, then other skaters that do not belong to G may overtake G to eventually win the race. This may be a main reason why somebody leads a group without a long stagnation. 

\begin{figure}[bt]
	\centering
	\includegraphics[width=4in]{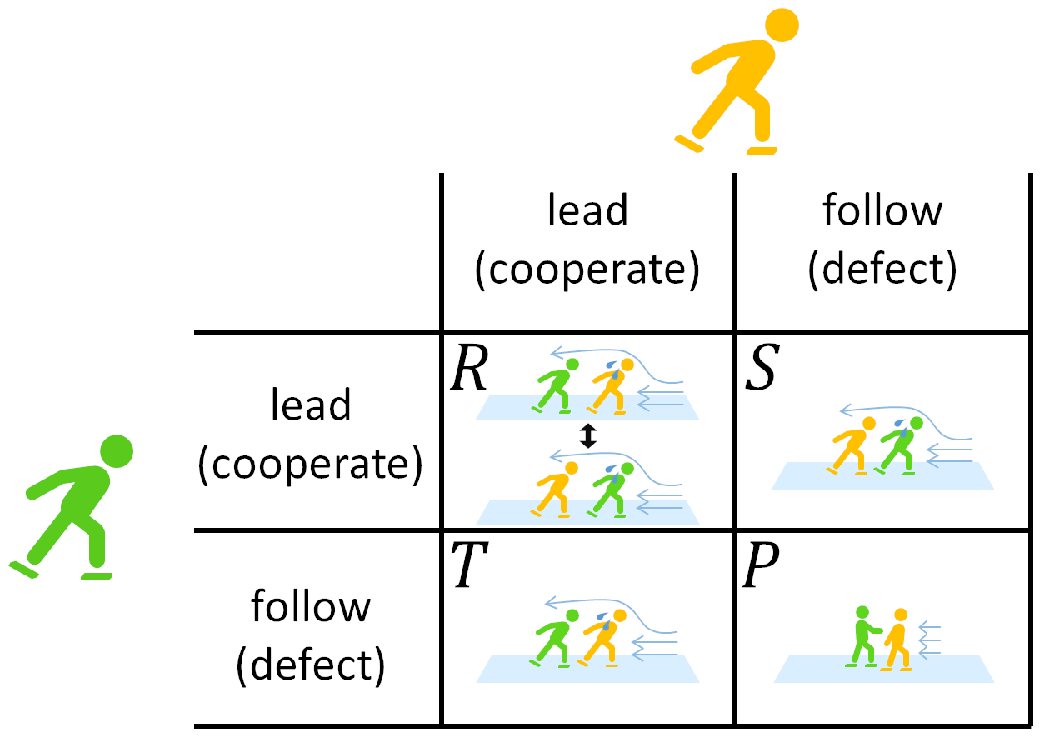}
	\caption{Schematic of the chicken game played by skaters, or the skater's dilemma game. The payoff values represent that for the row player.}
	\label{SkaterDilemma}
\end{figure}

If this is the case, one can liken the situation for the skaters to a multi-person variant of the chicken game, which is a social dilemma game. 
The chicken game for skaters, which we call the ``skater's dilemma'' is schematically shown in Fig.~\ref{SkaterDilemma}.
In the original chicken game, each of the two players has two options, either to cooperate or defect 
\cite{Sugden1986book, Fudenberg1991book, Osborne1994book, Nowak2006book}. In mass start races, cooperation corresponds to leading a group, and defection corresponds to hiding behind somebody to avoid air resistance. There are three situations to be distinguished. First, if one skater cooperates and the other skater defects, the defector obtains the highest payoff, denoted by $T$, corresponding to a high probability of winning the race. The cooperator gains a moderately low payoff, denoted by $S$, corresponding to a lowered probability to win the race. Second, if both players defect (i.e., neither skater wants to lead a group),
both skaters gain payoff $P$. If the other skater defects, it is better for a skater to cooperate (i.e., lead a group) than both players defecting because, if both skaters try to avoid leading the group, they may be easily superseded by other skaters. Therefore, the payoff under mutual defection, $P$, is considered to be smaller than $S$. Third, if both players cooperate (i.e., both skaters offer to lead a group), each of them gains payoff $R$. Payoff $R$ should be less than $T$ because mutual cooperation does not imply that a cooperative skater uses the other cooperative skater as a shield at least for the entire period. Payoff $R$ is probably larger than $S$ because mutual cooperation does not imply that a cooperative skater is not used as a shield of the other cooperative skater at least for the entire period. Therefore, we expect that $T > R > S > P$, which is the definition of the chicken game. A straightforward extension of the chicken game to multiperson situations is to assume that each player selects either cooperation or defection and plays the two-person chicken game with all the other players in the same group. For this game, the Nash equilibrium is a mixture of cooperators and defectors, where the fraction of cooperation is given by $(S-P)/(T-R+S-P)$ \cite{Colman1995book}. Therefore, the empirical skater's dilemma may correspond to the case in which $(S-P)/(T-R+S-P)$ is small because if one skater leads a linear group of skaters, all skaters but the leader can avoid high air resistance. However, a linear group of skaters implies that different defectors (i.e., non-leaders) are in fact not in the same situation. Positioning right behind the leader and positioning far down the line, for example, should differently impact the likelihood to be top finishers. Then, we probably need to consider more complicated games to realistically model skater's dilemma situations. Investigating mass start races and similar types of competition using the chicken or other games, combined with multi-agent modeling, also warrants future work. 

In team sports, athletes are often motivated to conduct social loafing, with which they avoid putting full effort and count on other team members' hard work. Social loafing occurs when the team members share a goal and are involved in the same task, such as in rowing and tug-of-war \cite{Latane1979JPersSocPsychol, Cashmore2008book}. We provided compelling evidence that skaters in mass start races are motivated to hide behind others to avoid energy consumption. However, this phenomenon is distinct from social loafing because skaters in a race do not form a team and they do not share a goal. They eventually have to beat other skaters. Therefore, except in some special cases such as when the skaters from the same country help each other in a race, the skaters in a race are involved in a zero-sum game. Individual skaters avoid full effort in the middle of the race as part of their strategic choice, rather than as social loafing.

The present work has a number of limitations. First, we had to impute data because we depended on publicly available video footages. Skaters often disappeared from the footage. Furthermore, we manually determined the times at which the individual skaters passed landmarks and the duration for which they are behind another skater. Because it is difficult to cover all the skaters by conventional videos that are mainly released for entertainment, it is desirable to have either videos being taken at the fixed location right above the arena to overview the entirety of the oval, or for skaters to carry RFID tag chips or wearable GPS trackers with which to record their locations \cite{Gudmundsson2017ACMComputSurv}. Then, one will be able to collect data more automatically and accurately.
Alternatively, in Tour de France, which is one of major cycling races, NTT Ltd.~has been acting as a technical partner and providing real-time data of the races since 2015 \cite{NTT-ebook}. Every rider attaches a GPS tracker underneath his saddle, which generates real-time data including the rider's speed, distances between riders, and their relative positions within the peloton. Anybody has access to the data in real time.
Submitting this or similar data set to the analysis pipeline proposed in this study may be an interesting exercise.
Second, we analyzed a relatively small number of races. Once the data collection is automated and as there are more races happening, we should be able to carry out an analysis on a much larger scale. Then, we may be able to more readily include other strategic factors such as the height of skaters (cf., shorter skaters are easier to hide behind others) and nationality (cf., skaters from the same nation tend to help each other during races) as variables. Despite these and other limitations, we believe that the present work provides empirical support of the role of a strategic behavior of skaters; they may be skating under a social dilemma game.

\section*{Acknowledgments}
We thank Takashi Kawakami and Masahiro Yoshida for technical advice on mass start speed skating.
G.I. thanks Mutsuyo Suzuki for the coordination between the skating club in Yamanashi Gakuin University and the authors.

\end{document}